\documentclass[letterpaper]{article}

\usepackage{graphicx}
\usepackage{iccc}
\usepackage{times}
\usepackage{helvet}
\usepackage{courier}
\usepackage[colorinlistoftodos]{todonotes}
\usepackage{subcaption}
\usepackage{wrapfig}
\usepackage{url}
\usepackage{tikz}
\usepackage{verbatim}
\title{Explainable Computational Creativity}

\author{Maria Teresa Llano\\
SensiLab, Monash University \\
Melbourne, Australia \And
Mark d'Inverno\\
Goldsmiths, University of London \\
London, United Kingdom \And
Matthew Yee-King\\
Goldsmiths, University of London \\
London, United Kingdom \AND
Jon McCormack\\
SensiLab, Monash University \\
Melbourne, Australia \And
Alon Ilsar\\
SensiLab, Monash University \\
Melbourne, Australia \And
Alison Pease\\
Dundee University \\
Dundee, United Kingdom \And
Simon Colton \\
SensiLab, Monash University \\
Queen Mary, University of London
}

\begin{document} 
\maketitle
\begin{abstract}
Human collaboration with systems within the Computational Creativity (CC) field is often restricted to shallow interactions, where the creative processes, of systems and humans alike, are carried out in isolation, without any (or little) intervention from the user, and without any discussion about how the unfolding decisions are taking place. Fruitful co-creation requires a sustained ongoing interaction that can include discussions of ideas, comparisons to previous/other works, incremental improvements and revisions, etc. For these interactions, communication is an intrinsic factor. This means giving a voice to CC systems and enabling two-way communication channels between them and their users so that they can: {\em explain} their processes and decisions, {\em support} their ideas so that these are given serious consideration by their creative collaborators, and {\em learn} from these discussions to further improve their creative processes. For this, we propose a set of design principles for CC systems that aim at supporting greater co-creation and collaboration with their human collaborators. 
\end{abstract}

\section{Introduction}
Although systems from the field of Computational Creativity (which we will refer to from now on as CC systems) -- and more generally AI systems -- have been successful in different application domains,
a common challenge for users and researchers is that most of these systems behave as black boxes, limited to opaque interactions where processes and reasoning are completely unknown or obscure \cite{scherer:2016}. As a result, users are left questioning the nature of the system's decisions, or are discouraged to the notion of collaborating with them. These limitations have raised the need for the development of models that offer more clarity and transparency in order to improve the potential for human-machine interactions with CC systems \cite{muggleton:2018,bryson:2017}. 

The field of \emph{Explainable AI} (XAI) has grown in recent years with the goal of making black box systems more transparent and accountable through models of explanation that communicate the way decisions have been reached. Current approaches to XAI, which are commonly associated with popular but opaque machine learning methods, centre on providing explanations as part of the output of the system; i.e. the focus is on delivering a final result to a user alongside a rationale of how this result was created. However, the creative process is often performed in isolation, with no place for intermediate explanations as the process progresses, let alone place for exchanges of information that can exploit human-machine co-creation.

In this paper we propose {\em Explainable Computational Creativity} (XCC) as a subfield of XAI. The focus of this subfield is the study of {\em bidirectional} explainable models in the context of computational creativity -- where the term explainable is used with a broader sense to cover not only one shot-style explanations, but also for co-creative interventions that involve dialogue-style communications. More precisely, XCC investigates the design of CC systems that can {\em communicate} and {\em explain} their processes, decisions and ideas {\em throughout the creative process} in ways that are comprehensible to both humans and machines. The ultimate goal of XCC is to open up {\em two-way communication channels} between humans and CC systems in order to foster co-creation and improve the quality, depth and usefulness of collaborations between them.

Designing and implementing this type of communication is extremely complex; however, we believe it is important to start a discussion towards what is needed for more fruitful and productive partnerships between CC systems and their users. Creativity is not a solo act, it is a social activity that benefits from life experiences, from influences of people, and from the contributions of different collaborators. This is why we are interested here in co-creative CC systems and on addressing the lack of two-way channels of communication.
This may result in increased novelty, or higher-value creative collaborations, however it is difficult to anticipate how successful and in which way any creative collaboration can emerge, but we believe the kind of active collaborations proposed through XCC would ultimately enhance the human-machine collaboration experience and increase the engagement of users with CC systems.

%
%
%

In the rest of the paper, we first provide relevant background literature. Then we present an overview of the current state of the art of co-creation and explainable AI in general and in CC in particular. We follow by identifying design principles that we believe CC systems with explainable capabilities should have. We finish with a discussion outlining challenges that need to be considered, opportunities that may arise, conclusions and directions for future work. 

\section{Motivation and Related Work}
\label{sec:background}
\label{sec:background}
\begin{figure}
\begin{center}
\includegraphics[scale=0.4]{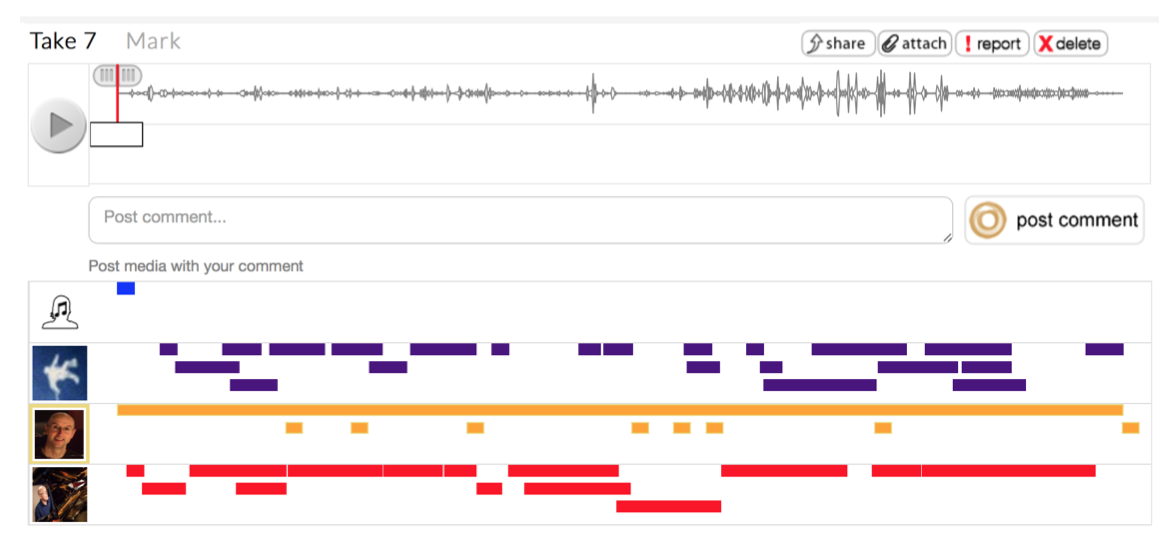}
\end{center}
\caption{The set of annotations created during the evaluation of SpeakeSystem. At the top of the image is the audio timeline. The coloured boxes below represent the annotations created by each of three annotators. Here the system cannot contribute to the conversation; i.e. the system does not have a voice.}
\label{fig:SpeakeSystem}
\end{figure}

Co-creation, real-time interactions and collaboration are important topics in the CC community; however, the communication between CC systems and their users is limited to a few exchanges, or the rationale behind their individual actions is often unknown by either, or there is little or no opportunity for discussion of the ideas presented, and many system outputs are discarded without a second thought. 

Take for instance the \emph{SpeakeSystem}: a real-time interactive music improviser which takes as its input an audio stream from a monophonic instrument, and produces as its output a sequence of musical note events which can be used to control a synthesizer. The BBC Radio 3 Jazz Line Up programme commissioned SpeakeSystem in 2015 for a live human and computer performance with alto saxophone player Martin Speake\footnote{https://www.bbc.co.uk/programmes/p033s4gj}\footnote{A recording of the performance and the source code for the system is available in an open source repository \cite{yeeking2016speake}}. Figure \ref{fig:SpeakeSystem} shows a recording of the performance that was uploaded to a timeline annotation system in order to carry out a focused discussion about it. The conversation involved the musician, the algorithm designer and a member of the audience. However, a key participant of the performance was not involved, the interactive music improviser. To illustrate, at some point of the recording the human musician commented:
\begin{quote}
\textit{``I think by this stage that I wanted the algorithm to come up with a new interaction mode.''}
\end{quote}
while an audience member said:
\begin{quote}
\textit{``I wonder what you are both thinking going into this section. The algorithm not a lot I suspect! Otherwise it would play notes.''}
\end{quote}
In a human-human interaction, the musician being `questioned' would reply with a rationale about his/her decision at those points in the performance; however, in this conversation it is not possible to know what the rationale or motivations of the system were as the system does not have a voice in the conversation. Instead the algorithm designer plays that role by explaining how the system works; i.e. how he designed the system. For instance:
\begin{quote}
{\bf Designer:} \textit{It has a kind of reset function that causes it to forget your patterns occasionally.''} \\
{\bf Human musician:} \textit{Yes I did wonder as sometimes it seemed to have logic in how it responded and then at other times it didn't make sense to me.''}
\end{quote}
Even though the algorithm designer provides information about how the system operates, he is unable to provide an explanation of what was going on at specific points in the performance as only the system ``knows'' the details of what went on at every point of the performance.

If the SpeakeSystem had been equipped with communication capabilities, not only it could have provided insights about its creative process in the conversation above, but the human musician could have communicated his intentions during the performance (perhaps through some kind of visual or haptic signal), giving the system a chance to respond during co-creation. This could have resulted in a more engaging experience, for the musician and the audience alike. 

\subsection{Co-creation, Real-Time Interactions and Collaboration in CC}
Models to improve collaboration and co-creation within CC have been explored in the community. In \cite{davis:2015}, for instance, an enactive model of creativity is proposed for collaboration and co-creation. The key principle for this model is that the system has a level of awareness and that co-creation happens through a real-time and improvised interaction with the environment and other agents. Another approach is mixed-initiative co-creativity \cite{yannakakis:fdg14}, which exploits a bi-directional communication based on the collective exploration of the design space and human lateral decisions that are used by the system to guide the creative task. Although these models establish communication channels, these are limited to an action/reaction type of model; i.e. there is little opportunity for further introspection in addition that they do not enable the systems to further support their contributions.

The {\em You Can’t Know my Mind} installation of The Painting Fool \cite{colton:iccc14} presented real-time interactions with users that resembled some of the aspects we cover in this paper: the system provides explanations about its process, motivations, etc. (e.g. by providing a commentary alongside its output), the user provides content to guide the creative process (e.g. by expressing a particular emotion to be the inspiration for the painting), and the system utilizes visual cues to reveal what is going on (e.g. with the use of an on-screen hand while it paints the picture). Although these interactions simulate communication, it is mostly a one-way creative endeavour. 

The Beyond The Fence musical is also an important sample of human-machine collaboration in the field. The musical writers that were involved in this effort were enthusiastic about the possibilities of collaboration with CC systems; however, they highlighted some of the flaws they faced when working with them:

\begin{quote}
``\textit{We waded through probably a thousand pages of computer generated tunes to find the fragments and phrases that felt right for the show's needs}''  Musical writers comment in \cite{coltonBTF:iccc16}
\end{quote}

\begin{quote}
``\textit{Collaborating with computers is utterly unlike anything either of us have encountered before, and at times, it has been incredibly frustrating}'' Musical writers comment in \cite{coltonBTF:iccc16}
\end{quote}

These comments highlight that ultimately, for a long-standing (working) human-machine relationship to be sustainable in the context of creativity, there is a need for mechanisms that enable a more active partnership. As the musical writers put it:

\begin{quote}
``\textit{I rather think that the future holds ways of allowing human artists to work with computers more comfortably, and with more control of their output, ultimately to support and perhaps shape their own creativity in ways they might not have been able to envisage}'' Musical writers comment in \cite{coltonBTF:iccc16}
\end{quote}

\subsection{The need for communication}
Nickerson et al. \cite{nickerson:1968} described the increasing complexity of human-computer interactions based on the ability to communicate with one another ``the thing that, above all others, makes the man-computer interaction different from the interaction that occurs in other man-machine systems is the fact that the former has the nature of a dialogue''. This thought has been echoed by other researchers, who have emphasised that computers are `comparable' to humans in some dimensions when seen as collaborators, particularly as dialogue-partners \cite{kammersgaard:1988}. 

With the rapid growth of AI techniques, this discussion has gradually highlighted the existence of a higher level of intelligence when contrasting the views of `interaction as tool use' and `interaction as dialogue' \cite{hornbaek:2017}, arguing that utility and usefulness are the main aspects for the first type of interaction to work, while having a constant, simple, direct and natural communication and understanding between human and computer, is key for the second type of interaction to work. 

This distinction of interacting with a machine considered as a creative intelligence rather than a tool is a key motivation for this work. For this, we need to consider a broader set of aspects of human interaction that are otherwise ignored in the narrower view of systems as tools. For this broader view, exposing the creative process is crucial. Specifically we hypothesise that establishing \emph{two-way communication channels}, within a co-creative human-machine partnership, where the creative process is transparent as well as discussed, would improve interactions, build up trust on CC systems and encourage human engagement. 

\section{Overview on the state of the art of XAI in CC}
In \cite{zhu:2018} the authors defined a new area of XAI which they called Explainable AI for Designers (XAID). The objective of XAID is to support games designers in specific design tasks. In their work they identified three spectra that describe co-creation in the setting of games design: i) spectrum of explainability, which ranges from understanding of the underlying operation of AI techniques to understanding of the input-output pattern, ii) spectrum of initiative, which refers to the level of intervention of the system (ranging from a passive tool to an active collaborator), and iii) spectrum of domain overlap, which is concerned with the degree of co-creativity that is needed (defined in terms of overlap of shared tasks). 

Relevant to this discussion is also the concept of framing as proposed in \cite{colton:iccc19}. Framing, as has been applied to date within the CC community, is mostly intended as a `final interaction'; i.e. to accompany an output with the expectation that it will increase its perceptive value. However, in \cite{colton:iccc19} the authors propose advocacy and argumentation as a potential purpose for framing. 

Both of the works mentioned above are relevant to our work and are a step towards the vision of CC systems playing a more active role in creative collaborations; however, XCC focus spreads not only to co-creation but also to other interactions that occur when producing a creative act; such as setting up an initial goal, delivering the product to a final user, producing feedback, etc. Moreover, we ground the interventions of a CC system not only on the current act of creation in which it is involved, but also in past experiences (i.e. we propose that CC systems should have a memory of their work). Additionally, we also adopt the notion of argumentation and advocacy as a role for XCC; but our model, proposes this role not only as a way to support a creative artefact, but also as a way to increase their involvement since the conception of an idea towards the production of it.

In the next section we outline some design principles for systems with XCC capabilities. We use a running example to illustrate our ideas using linguistic communication as the primary medium for explainability; however, we consider communication in its broader sense, not just through linguistic forms. We will expand on this later in the paper.

\section{Design Principles for XCC Systems}
\label{sec:approach}
The main objective of the design principles outlined next is to enable both CC systems and their human users to communicate with each other so that there is a common and clear understanding throughout their interactions. We have drawn from a range of research that looks at human collaboration, teamwork, cognitive science and psychology, as well as from our experience on the development of computational creative systems. From this, we have identified four main design principles: \emph{mental models, long-term memory, argumentation} and \emph{exposing the creative process}. We now explain each principle in detail.

\subsubsection{Mental models:} 
Are representations of key elements of the creative environment that help conceptualize, understand and construct expectations of how things work and how individuals interact within a creative collaboration \cite{mccormack:2020,mohammed:2010}. The concept of a mental model comes from psychology and cognitive science research \cite{craik1967,JOHNSON-LAIRD93} and has more recently been successful in HCI \cite{Norman1982,Krug2014}. Having a good mental model of how software works is vital for it's usability.

{\em Relevance to CC:} The idea of mental models as a key aspect for the design of real-time co-creative systems has been highlighted previously in \cite{mccormack:2020}. Mental models consider a broad set of aspects of human interactions that would aid the understanding of essential elements within a collaboration and of the interactions throughout it. 
CC systems can use these representations, not only to understand the operation of their co-creators, the environment and the domain, but also to reason about them and seek different (creative) ways of interaction by questioning them; for instance by playing a bossa nova style chord progression when improvising with a musician with a strong rock background (i.e. challenging preferences). Shared mental models can increase CC systems' awareness of processes within the collaboration. For instance, in order to achieve cognitive convergence when the mental models of the collaborators do not match (e.g. when they are not moving towards the same goal) \cite{fuller:2010}.

Equipping CC systems with mental models, of both themselves and of their co-creators, not only would enable better coordination when trying to come up with something new, but would also provide a valuable resource for CC systems to explain, justify and defend their contributions. We believe this increases the capacity, of both CC systems and their users to generate appropriate and complementary output. As pointed out in  \cite{bruhlmann:2018}, people are more motivated to use a specific technology when it is congruent with their personal values, goals, and needs.

{\em Features:} Relevant elements of mental models would involve team aspects such as goals, roles, capabilities, expectations, etc., domain aspects such as conventions governing the operation of particular domains, stakeholders profiles, relationships between elements of the domain, etc., as well as interpersonal aspects of individuals such as preferences and how they communicate. Sharing the underlying representation of a mental model would depend highly on the domain at hand. The dynamics of collaborations in different domains influence the way mental models are built and used. For instance, improvisational settings are governed by implicit interactions with subtle signs and cues, while in an advertising campaign setting, for instance, participants can explicitly communicate their intentions, discuss their ideas and establish agreements. 

{\em Example:} Imagine for instance the following interaction between an Advertising Executive (AE) and an XCC system (XCCS) when collaborating to design an advert for a toothpaste (goal) in which the clients would like to emphasise the ideas of teeth and decay (domain information):

\begin{figure}
  \centering
    \includegraphics[width=0.3\textwidth]{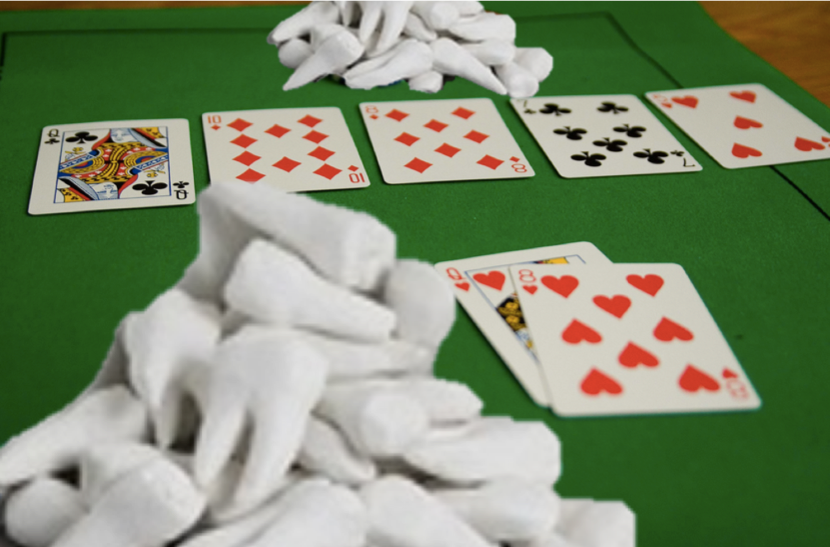}
  \caption{First attempt.}
  \label{fig:poker}
\end{figure}

\noindent \textbf{XCCS}: How about the image in Fig~\ref{fig:poker}?
Teeth and dice are similar, as they are white, cube-shaped and shiny.
Dice connects to gambling, which connects to poker, so I got the idea: {``\em What if someone gambled with
their teeth instead of with money?''} \\
\noindent \textbf{AE:} I don't like it, because it's too big of a
jump -- the connection to toothpaste is not obvious.

The system can pick up from the explicit intervention that the AE does not like ideas whose connections are not obvious and can adjust its mental model with this information in order to generate new ideas.  

\subsubsection{Long-term memory:} 
Is the capacity of storing and accessing information of past experiences and interactions. 

{\em Relevance to CC:} Studies in cognitive psychology have found that memory is a crucial element of creativity and that an important part of the creative process happens by drawing on previous experiences and the information we have in our memories \cite{madore:2015}. Being able to explain their decisions and support its ideas requires CC systems to have a memory of the processes, decisions and interactions they have undertaken in current and previous collaborations. 
``The act of remembering is an attempt to recreate events and experiences that have occurred in the past'' \cite{stein:1989}. Equipping CC systems with this ability would enhance the creative capacity of their collaborations. Possessing a memory would also serve other purposes such as breaking habits and avoiding repetition or mistakes. 

{\em Features:} Enabling a bi-directional communication would provide opportunities for CC systems to store different types of information, such as failed attempts, successful artefacts, strategies used, temporal information, users' reactions, etc. How to store and access this information is an important aspect to consider here. For instance, one methodology could be as defined in \cite{davis:2015}, where the authors use the concept of {\em perceptual logic} to classify information in a way that aids co-creation as follows: in local perceptual logic the system only considers specific details (such as a line in a drawing), in the regional perceptual logic the information is grouped into clusters (e.g. straight lines, lines that are close to each other, etc.), while in the global perceptual logic, the system considers relationships between regions (e.g. identifying that the left hand side of a drawing has fewer lines than the right hand side). Depending on the domain and the purpose of a CC system in that domain, different mechanisms may use to handle such memory. 

{\em Example:} Let us take for instance the example of the AE and the XCCS working on the toothpaste campaign. The AE did not like the idea for the ad even after the XCCS provided an explanation. The system can then review past experiences and find additional strategies to support its idea, in doing so the system finds that in a previous interaction another AE used a tag line to clarify an abstract concept for an ad. The interaction then follows:

\noindent \textbf{XCCS:} I think the connection is good. There is a strong {\em
  surface similarity} between teeth and dice and the idea is {\em surprising}
as the two concepts are not normally associated; maybe we could add a tagline to support the
image? Something related to the not apparent connection between teeth, gambling and money? \\
\noindent \textbf{AE} That is good. I like the idea of using a tagline. Maybe something like {\em ``There are some things you just can't afford to gamble with''}. \\
\noindent \textbf{AE:} There may be something here, but I'm not sure yet...


\subsubsection{Argumentation:} Is the process of reasoning-about and supporting specific contributions within a creative collaboration.
 
{\em Relevance to CC:} A fruitful creative partnership enables participants to both explain and justify their ideas in order to unveil or clarify their creative value. An artefact may be of poor quality but the process behind it novel and interesting. CC systems that cannot further support and champion their creative contributions carry the risk of being prematurely discarded. 

{\em Features:} As pointed out in \cite{colton:iccc19}, argumentation provides a set of very valuable resources that can be used by CC systems to enhance communication with their users. Take for instance the theory of critical questions \cite{2008:argumentationSchemes}, which helps anticipate questions or concerns that may arise in specific situations with different stakeholders, as well as the mechanisms to try and address possible conflicts that these questions or concerns may arise within a creative collaboration. 

Ultimately this part of the process involves an attribution or cognitive process that requires different dimensions to be taken into account. Is it a one-shot argument, or is it the start of a dialogue? Is the argument targeted for a co-creator, an end user, a member of the audience? What elements of the mental model should most influence the selection of arguments? When is the right moment to push or to stop? What kind of language, visualisation or medium should be used to convey the argument? What does the system want to achieve with it: clarification, persuasion, feedback, etc.?

{\em Example:} Take for instance the example of the AE and the XCC system in the toothpaste advert campaign. After the second interaction, when the system proposes the use of a tagline to clarify the image, the AE is willing to give a chance to the system's idea. A further interaction could go this way: 

\noindent \textbf{XCCS:} The idea and image are somewhat {\em repulsive} and repulsive adverts have a shock value which help people remember them. Other repulsive adverts have been quite effective. Have a look at this: \url{https://www.youtube.com/watch?v=AOph5V78oxs} (video showing an advert warning about the addictive nature of cocaine).\\
\noindent \textbf{AE:} Hmmm... possibly.  

Without the system pushing back a little, it is easy to imagine that a user may stop engaging after the first generative act, thus missing out on the concept.

\subsubsection{Exposing the creative process:} consists of opening up the environment and exposing the steps, assessments, metrics, influences, etc. that constitute the processes and decisions within the operation of a domain.

{\em Relevance to CC:} Providing an explanation for a process or a decision is a useful way to obtain a better understanding of what is going on within a closed environment; however, descriptions or clues envisaged for this purpose may sometimes fall short in aiding that understanding. Opening up the environment and exposing the steps, assessments, influences, etc. that constitute those processes and decisions would provide a clearer view of how a CC system works. We believe that observing, feeling, or in some way sensing the underlying structures of a CC system, instead of being told how things work, may trigger thought processes in the mind of their co-creators that may result in benefits for the creative collaboration.

{\em Features:} Exposing the creative process requires the models and mechanisms used by a system to be {\em interpretable} so that they can be easily and unambiguously communicated to others. The development of interpretable models is being encouraged in the AI community because of the inherent problems with unexplainable models, such as unfaithful accounts of their computations \cite{rudin:2019}. In the absence of interpretable models, CC systems should be equipped with interfaces that make communication with their users interpretable.  In other words, the way a CC system communicates must be simple and precise, yet the communication needs to be meaningful so that the new information helps progress the collaboration. An example of this is provided in \cite{mccormack:CHI19} where the authors equip an AI musician (whose underlying model consists of a neural network) with the ability to continuously communicate how confident it feels during an improvised performance. Human performers also implicitly communicated their confidence to the computer via biometric signals. The work showed that this type of simple, interpretable, communication increased the flow within the human-AI collaboration and the quality of the music produced.

{\em Example:} Let us take for instance the example of the AE and the XCC system working on the toothpaste campaign. The AE is still not sure about the idea for the ad. The system can then expose its reasoning even further so that the AE has a better understanding where the ideas come from. Imagine for instance the CC system has an interface that allows the user to investigate the underlying structures behind its ideas through the mean of a visual graph representation of the system's knowledge base.
%

\begin{figure}
  \centering
    \includegraphics[width=0.3\textwidth]{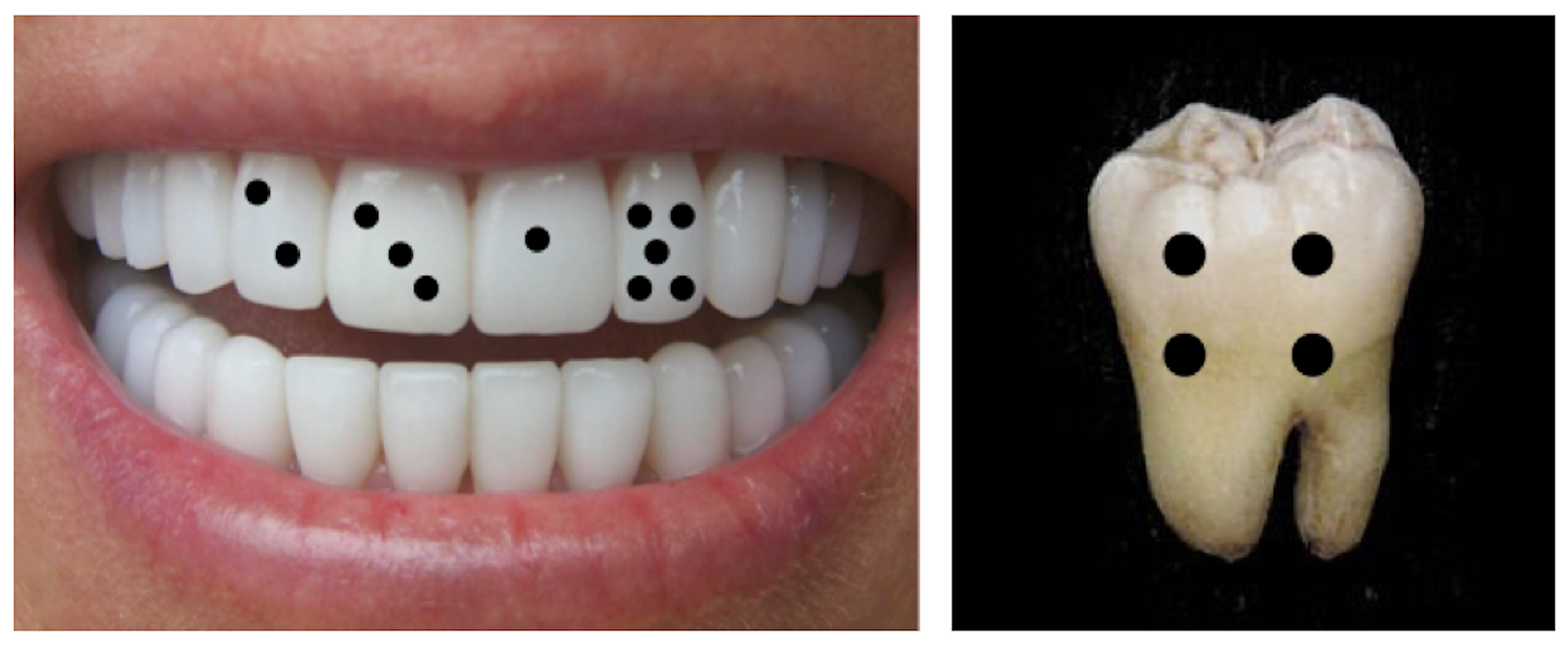}
  \caption{Second attempt.}
  \label{fig:first-blend}
\end{figure}

\begin{figure}
  \centering
    \includegraphics[width=0.2\textwidth]{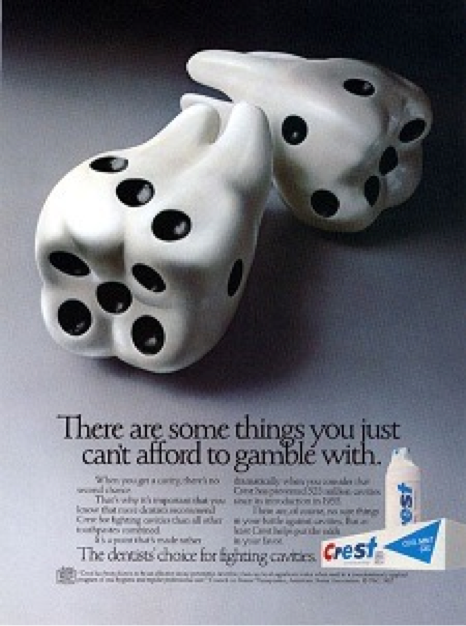}
  \caption[Final result.]{Final result.\footnotemark}
  \label{fig:final}
\end{figure}

By seeing the connections among the concepts in the knowledge base, the user realises that the two related concepts; i.e. dice and teeth, are not actually directly connected and sees this as an opportunity to improve the concept of the ad. The AE proceeds as follows:

\noindent \textbf{AE:} I still don't like the image. How about {\em blending} the two similar concepts: dice and teeth? \\
\noindent \textbf{XCCS:} Ok. I've made the images in
Fig~\ref{fig:first-blend}. I prefer the first one, as the cube shape
is clearer. \\ 
\noindent \textbf{AE:} I prefer that one too. I'll adapt it and
add tagline, brand and product information... Here is the final
advert, in Fig~\ref{fig:final}.

\footnotetext{Original advert taken from
  \url{http://bit.ly/2uucLqR}}

\subsection{A framework for XCC}
\begin{figure*}
\begin{center}
\includegraphics[scale=0.5]{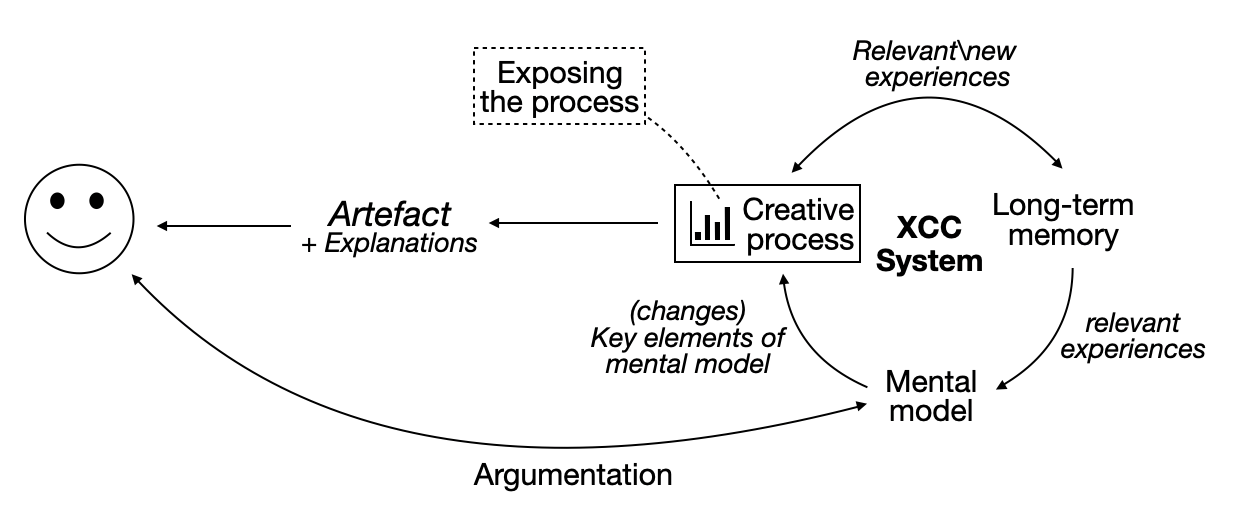}
\caption{XCC design principles: the creative process is shaped by relevant past experiences and the shared mental model, which is itself shaped by the arguments exchanged with the user as well as by relevant past experiences. The long-term memory is updated with new experiences of the creative process, and the operation of the system is exposed through different interfaces that allow the user to understand underlying procedures of how the system works.
}
\label{approach}
\end{center}
\end{figure*}

Figure~\ref{approach} summarises the kind of interactions of the proposed approach based on the four principles presented in this paper. Instantiating these interactions naturally depends on the creative domain, the stakeholders involved, and the stage of the creative process where the interventions occur. 

The domain is the most important factor to determine the kind of medium for communication. In music, for instance, communication may most likely occur through the music itself -- imagine for instance a CC musical composer demonstrating through a virtual keyboard how a pianist should emphasise a particular phrase --, while in painting this communication may occur through brush strokes -- imagine now a CC painter-collaborator that paints all over a section of a painting they think should be emphasised \cite{mccormack:2020}.

The stakeholders greatly influence the sources and type of information that these interactions would require. A co-creator may need technical details of the operation of a system -- an ideation system may provide a tree representation of the relevant concepts from the knowledge base from which an idea was produced (as in the advertising example), while for an end-user the intuition would probably be more useful -- for instance a CC poetry composer that explains the mood reflected in the poem because it read sad news in the newspaper (the end-user does not need to know, for instance, the technical mechanism of sentiment analysis used). 

The stage of the creative process is most influential on how the elements of the design principles are managed. In a preparatory stage, when the collaboration is just starting, the interactions between CC systems and their users help set up the context of what the collaboration is about; i.e. a shared mental model is agreed -- take for instance the motivating example of the SpeakeSystem: this stage would have allowed the system to inform the human musician about its reset function, which could have avoided the human musician wondering why the system {\em didn't make sense} at some points during the performance. In a co-creation stage the interactions require an
iterative process that involves a constant revision of the mental model in order to ensure that the collaboration is converging towards the same goal, the addition of new memories (which reflect the experience of the current interactions), access to old memories, and possibly various cycles of generative acts -- here for instance, the human musician working with the SpeakeSystem could have signalled during the performance (through a visual cue or facial expression) when he wanted the system to challenge him instead of only responding to him. Finally, a post-creative stage provides an outlet to present the artefact as well as for feedback and reflexion. Such an outlet may also represent an opportunity for the revision of the mental models and for adding new memories to the system -- here for instance the SpeakeSystem could have provided an explanation to the member of the audience who wanted to know what was going on at a certain point of the performance:
\begin{quote}
\noindent {\bf Human annotator:} ``{\em I wonder what you are both thinking going into this section. The algorithm not a lot I suspect! Otherwise it would play notes}''\\
\noindent {\bf System:} {\em I was enjoying what my partner was playing here. He/she seemed to really like this piece. He was very confident, was ``in the groove}''
\end{quote}

As previously mentioned, we consider communication in its broadest sense, not just through linguistic forms. In general, the form of communication should be appropriate to the creative task and information being communicated. For example, in an improvised music performance, it would be inappropriate for participants to stop playing and start talking about why one made a particular musical decision in the previous bar. Often visual, sonic, haptic or even olfactory communication may be the most suitable, and the form of communication can influence the creative outcomes in subtle, non-obvious ways. For example, the smell of freshly baked bread or a spring meadow can trigger specific memories or support synaesthetic-like relationships to other sensory media \cite{ackerman1990}.

\section{Discussion}
\subsection{Challenges and Opportunities}
Different issues surround the idea of explainable models. A concern that has been raised in the XAI field is that explainable models are not accurate representations of the actual functionality of the system in some aspects of the feature space. According to \cite{rudin:2019} ``an inaccurate (low-fidelity) explanation model limits trust in the explanation, and by extension, trust in the black box that it is trying to explain''. 
Although we believe that CC provides an outlet in which explanations do not have to be faithful to the intrinsic motivations/objectives of a system (as has been postulated through the notion of framing in CC), we need to be careful so as not to endanger the trust that co-creators, users and audiences impose into these systems. For instance, the CC advertiser in our example may be involved in pitching the concept of the ad to the client. In the process it may provide intuitions behind the campaign but in doing so it should not take credit for aspects of it that were actually ideas of its co-creators (e.g. the idea of blending dice and teeth in the example). Another common challenge is that explainable systems have a social responsibility. This has been explored extensively in \cite{arrieta:xai20} where the concept of Responsible AI is introduced together with some practical principles (such as fairness, transparency, and privacy). For instance, one thing is for the CC advertiser in our example to refer to gambling as an analogy to endangered oral health, and another (questionable possibility) is to use it to encourage gambling. Finally, there must be clear limits about the extent of which a CC system may try to champion an idea; i.e. CC systems cannot disrupt or take control over a creative process. Imagine for instance the CC advertiser in our example not backing down about using the first image of the poker table even tough its co-creator had already rejected it. 

Despite these challenges, explainable models, as suggested here, may open up possibilities of new kinds of social interactions between CC systems and their users. Work from the social sciences has shown that non-human entities can take up active roles in social practices and that imposing concrete boundaries or definitions between what roles these can or cannot play only limits their potential. Strengers \cite{strengers:2019} illustrates this point by looking at {\em roomba riding}, a trend that refers to how pets enjoy `riding' a robotic vacuum cleaner called the Roomba, and the potential of this type of technology to become a pet entertainment device. Imagine for instance a CC collaborator that strongly suggests its human co-creator to stop working during the weekend to be with his/her family, or a human co-creator that opens up a co-creative collaboration with a CC system on social media, or a CC system that proposes its human co-creator to watch a film together in order to have a break and maybe get some inspiration. Through this, the idea is to emphasize that in order to identify potential relational roles and dynamics between humans and machines, it is important to not only consider humans as capable of performing a social role, but machines should also be considered as capable performers of such roles. Equipping CC systems with explainable capabilities may reveal a type of dynamics between humans and CC systems that is not possible with most current technologies. 

\subsection{Conclusions and Future Work}
We have presented Explainable Computational Creativity (XCC), as a sub-field of XAI that is focused on the applicability of explainable models in the area of CC and how these can be used to open up bidirectional communication channels between CC systems and their users. We have outlined four design principles we believe are crucial for these types of models, namely mental models (individual representations of how things work), long-term memory (the capacity to store and access details of past experiences), argumentation (the ability to reason and support creative contributions), and exposing the creative process (revealing specific details about the operation of a system). We believe that this framework will benefit human-CC interactions by: i) enhancing their creativity (as a result of working together rather than in isolation), ii) creating more fruitful and productive partnerships (by establishing bi-directional communication channels where creative contributions can be examined), and iii) increasing the engagement of users with CC systems (by improving the flow of these collaborations and the overall perception of their outcomes).

The proposed approach opens up questions such as: How can CC systems explain their creative acts and how do people respond to them? What makes a good or poor explanation in the context of CC? What are the challenges that CC systems face when providing explanations? How can CC systems change their behaviour in response to feedback, new ideas or historical knowledge of the co-creators they are working with? How do we evaluate CC systems with explanatory capabilities? We have explored the surface of some of these questions but as mentioned before, domain specific instantiations would dictate different interactions. We plan to explore this as future work in the context of music improvisation and composition, by extending the SpeakeSystem with explanatory capabilities.

\bibliographystyle{iccc}
\bibliography{iccc,refTeresa,teresa,MyReferences}

\end{document}